\documentclass[cpp,fleqn]{w-art}
\usepackage{times}
\usepackage{w-thm}
\usepackage[utf8]{inputenc}
\usepackage{amsmath}
\usepackage{amssymb}
\usepackage{graphicx}
\usepackage{subfigure}
\usepackage[english]{babel}
\begin{document}
\DOIsuffix{theDOIsuffix}
%%
%% issueinfo for header and copyright line
\Volume{42}
\Issue{1}
\Month{01}
\Year{2013}
%%
%%    First and last pagenumber of the article. If the option
%%    'autolastpage' is set (default) the second argument may be left empty.
\pagespan{1}{}
%%
%%    Dates will be filled in by the publisher. The 'reviseddate' and
%%    'dateposted' (Published online) entry may be left empty.
\Receiveddate{}
\Reviseddate{}
\Accepteddate{}
\Dateposted{}

%% \pretitle{Editor's Choice}

%% We have a short and a long form for the title. The short form
%% (optional argument) goes into the running head.

\title[Hubbard nano-cluster dynamics]{Dynamics of Hubbard nano-clusters following strong excitation}

%% Please do not enter footnotes or \inst{}-notes into the optional
%% argument of the author command. The optional argument will go into
%% the header.  If there is only one address the marker \inst{x} may be
%% omitted.

%% Information for the first author.
\author[M. Bonitz]{M. Bonitz\footnote{Corresponding
     author: e-mail: {\sf bonitz@physik.uni-kiel.de}, Phone: +49\,431\,8804122,
     Fax: +49\,431\,8804094}\inst{1}}
\author[S. Hermanns]{S. Hermanns\inst{1}} 
\address[\inst{1}]{Christian-Albrechts-Universität zu Kiel, Institut für Theoretische Physik und Astrophysik, Leibnizstraße 15, 24098 Kiel, Germany}
%%
%%    Information for the second author
\author[K. Balzer]{K. Balzer\inst{2}}
\address[\inst{2}]{University of Hamburg, Max Planck Research Department for Structural Dynamics, Building 99 (CFEL), Luruper Chaussee 149, 22761 Hamburg, Germany}
%%
%%    Information for the third author
%%
\newcommand{\todo}[1]{\textcolor{red}{\underline{#1}}}
\newcommand{\eq}[1]{Eq.~(\ref{#1})}
\newcommand{\nn}{\nonumber}
\newcommand{\e}[1]{\mathrm{e}^{#1}}
\newcommand{\op}[1]{\hat{#1}}
\newcommand{\bra}[1]{\langle{#1}|}
\newcommand{\ket}[1]{|{#1}\rangle}
\newcommand{\braket}[2]{\langle{#1}|{#2}\rangle}
\newcommand{\eqsand}[2]{Eqs.~(\ref{#1}) and~(\ref{#2})}
\newcommand{\cc}{{\cal C}}
\newcommand{\mean}[1]{\langle #1 \rangle}
\newcommand{\tc}{T_{\cal C}}
\newcommand{\dc}{\delta_{\cal C}}
\newcommand{\ii}{\mathrm{i}}
\newcommand{\thc}{\theta_{\cal C}}
\newcommand{\intc}[1]{\int_{\cal C}\mathrm{d}{#1}\;}
\newcommand{\intlim}[3]{\int_{#1}^{#2}\mathrm{d}{#3}\;}
\newcommand{\mret}{{\mathrm{ret}}}
\newcommand{\madv}{{\mathrm{adv}}}

%%    \dedicatory{This is a dedicatory.}
\begin{abstract}
The Hubbard model is a prototype for strongly correlated electrons in condensed matter, for molecules and fermions or bosons in optical lattices. While the equilibrium properties of these systems have been studied in detail, the excitation and relaxation dynamics following a perturbation of the system are only poorly explored. Here, we present results for the dynamics of electrons following nonlinear strong excitation that are based on a nonequilibrium Green functions approach. We focus on small systems---``Hubbard nano-clusters''---that contain just a few particles where, in addition to the correlation effects, finite size effects and spatial inhomegeneity can be studied systematically.
\end{abstract}
%% maketitle must follow the abstract.
\maketitle                   % Produces the title.

%% If there is not enough space inside the running head
%% for all authors including the title you may provide
%% the leftmark in one of the following three forms:

%% \renewcommand{\leftmark}
%% {First Author: A Short Title}

%% \renewcommand{\leftmark}
%% {First Author and Second Author: A Short Title}

%% \renewcommand{\leftmark}
%% {First Author et al.: A Short Title}

%% \tableofcontents  % Produces the table of contents.
\section{Introduction}\label{s:intro}

Strongly correlated quantum systems and materials e.g.,~\cite{pavarini11} are of rapidly growing relevance in many fields of physics and chemistry. Especially the out-of-equilibrium dynamics are of great current interest in solid-state, atomic and molecular physics, in nanoelectronics, quantum transport etc.. In all these fields, the availability of intense and coherent radiation, combined with ultra-short laser pulses, has triggered many key experiments that allow one to investigate matter under extreme nonequilibrium conditions where strong correlations and nonlinear effects occur simultaneously \cite{balzer_jpcs13}. Examples are the photoionization of multi-electron atoms and molecules~\cite{becker12,schuette_prl12}, the many-body dynamics of particles in optical lattices~\cite{bloch08} or quantum interference effects in Mott insulators~\cite{wall11}.

From the theory side such systems pose particular challenges since quantum, spin and strong correlation effects have to be treated selfconsistently under situations far from the ground state or from thermodynamic equilibrium. Among the approaches that are capable to handle such problems we mention time-dependent density functional theory and density operator methods, e.g., \cite{book_bonitz_qkt,akbari_12,hermanns_jpcs13}. 
Recently, the nonequilibrium Green function (NEGF) approach has attracted particular attention. During the past 15 years it has been successfully applied to a variety of many-body systems in nonequilibrium, including the optical excitation of electron-hole plasmas in semiconductors~\cite{kwong98,kwong00}, nuclear collisions~\cite{rios11}, dynamics of laser plasmas \cite{bonitz_cpp99,haberland_pre01} and the problem of baryogenesis in cosmology~\cite{garny11}. More recently, NEGF methods have also been applied to finite spatially inhomogeneous systems, including 
the carrier dynamics and carrier-phonon interaction in quantum dots and quantum wells ~\cite{gartner06,lorke06,bonitz_prb07,balzer_prb09}, molecular transport in contact with leads, e.g.,~\cite{uimonen11,khosravi12} or small atoms, e.g.,~\cite{dahlen07,balzer_pra10,balzer_pra10_2}. For a recent overview on NEGF applications to inhomogeneous systems, see Ref. \cite{balzer_lnp13}.

Applications of NEGF methods to small Hubbard clusters have been presented not long ago \cite{puigvonfriesen09,puigvonfriesen10} and showed the great potential of this method. The physical features that could be explored include the relaxation dynamics, the excitation spectrum and, in particular, the relevance of double excitations \cite{balzer12_epl,sakkinen12}. At the same time NEGF simulations exhibited serious conceptual problems that are related to unphysical damping effects \cite{puigvonfriesen09} and computational difficulties that limited the spectral resolution and the duration of the  nonequilibrium propagation. We could recently solve (or at least, substantially weaken) both problems by invoking the generalized Kadanoff-Baym ansatz (GKBA) of Lipavski~\textit{et al.}~\cite{lipavsky86}. In order to test the quality of the second Born GKBA approach we concentrated on a one-dimensional (1D) Hubbard cluster containing just two sites and two electrons because here comparisons with available exact diagonalization methods are possible, cf.~\cite{hermanns_jpcs13,hermanns12_pysscripta}.
The goal of the present manuscript is to briefly discuss these results and extend them to larger systems as well as to 2D Hubbard clusters and to further discuss the capabilities of the NEGF approach in application to Hubbard nano-clusters.

\section{The Hubbard model and its nonlinear excitation dynamics}\label{s:hubbard}
As outlined above, we are interested in the dynamical behavior of a finite quantum system beyond the regime of linear response. To this end, we consider a  Hubbard model at half-filling with hopping amplitude $T$ and on-site interaction $U$.
The initial hamiltonian, for times $t<0$, reads
\begin{align}
\label{h0}
 \op{H}_{0}=-T\sum_{<s,s'>}\sum_{\sigma=\uparrow,\downarrow}\op{c}^\dagger_{s,\sigma}\op{c}_{s',\sigma}+U\sum_{s}\op{n}_{s,\uparrow}\,\op{n}_{s,\downarrow}\;,
\end{align}
where $s$ and $s'$ label the discrete sites, and $<\!s,s'\!>$ indicates nearest-neighbor sites. Further, $\op{n}_{s,\sigma}=\op{c}^\dagger_{s,\sigma}\op{c}_{s,\sigma}$ denotes the density operator, and the energy (time) is measured in units of $T$ (the inverse hopping rate $T^{-1}$). Generally, we study the chain for open boundary conditions but this is irrelevant for the special case of two sites.

At time $t \ge 0$ the system is strongly perturbed by an instantaneous change of the energy of site ``0''~\cite{puigvonfriesen09, balzer_jpcs13}, which leads to the perturbation,
\begin{align}
\label{h1}
 \op{H}_1=\epsilon_0\,\theta(t)\sum_{\sigma=\uparrow,\downarrow}\op{n}_{0,\sigma}\;.
\end{align}
\begin{figure}[t]
\centering
\subfigure{
\includegraphics[width=0.50\textwidth]{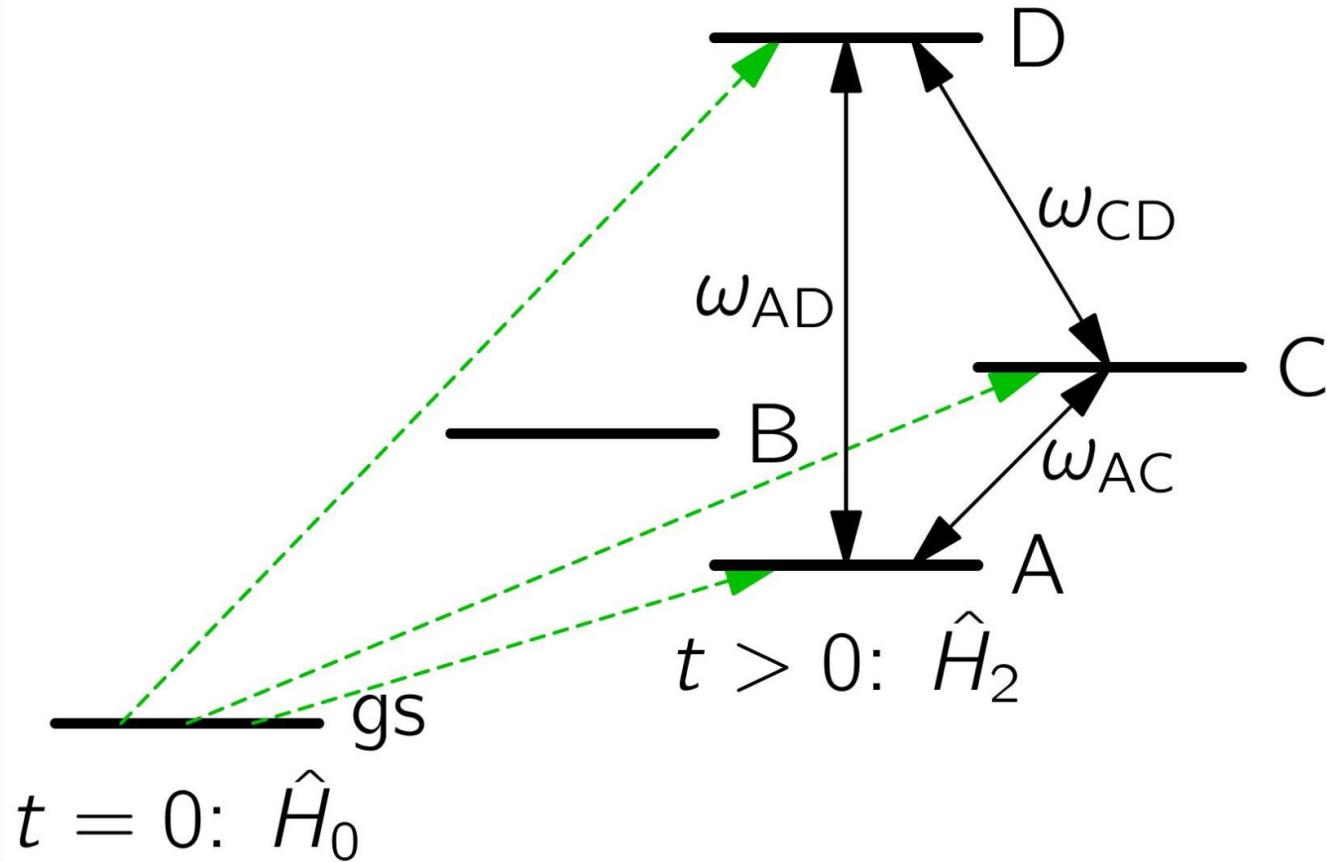}}
\hspace{1cm}
\subfigure{
\includegraphics[width=0.40\textwidth, height=5cm]{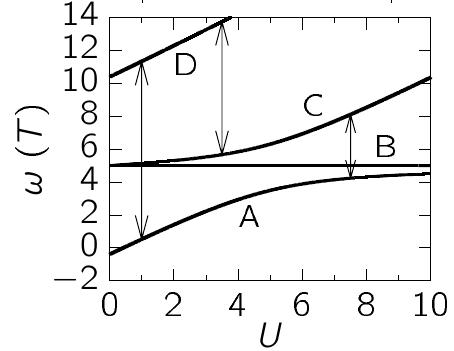}}
\caption{{\bf Left}: Schematic view of the excitation dynamics of two electrons in the two-site Hubbard system, starting from the ground state (gs) of hamiltonian $\op{H}_0$. For $t>0$, we switch to the new hamiltonian $\op{H}_2=\op{H}_0+\op{H}_1$, cf.~\eqsand{h0}{h1}. {\bf Right}: The four energies of the perturbed hamiltonian versus interaction parameter $U$.}
 \label{fig1}
\end{figure}
\hspace*{-0.12cm}The resulting change of the spectrum is sketched in Fig.~\ref{fig1} for the simplest case of two electrons in a two-site Hubbard cluster.
For $t>0$ and arbitrary real values of the parameter $\epsilon_0$, the perturbed system $\op{H}_0+\op{H}_1$ will initially show a depopulation of the  site $0$ followed by an accumulation of density on the second. Subsequently, also the density on the remaining site(s) will change with time and, finally, all occupations will start to oscillate. In the case of $\epsilon_0\ll 1$, the population change is expected to be small, such that the dynamics should be well characterized by the linear response properties of the chain. For $\epsilon_0 \gtrsim1$, however, we expect nonlinear effects to become crucial. 

To verify these qualitative predictions for the time dynamics, a computational study is indispensable. As a reliable theoretical tool we will use nonequilibrium Green functions theory which we briefly outline in the following section.

\section{Nonequilibrium Green Functions}\label{s:negf}

To describe the electron dynamics of the Hubbard nano-cluster following the rapid change of the hamiltonian, the central quantity is the one-particle nonequilibrium Green function defined on the complex Keldysh contour $\cc$, e.g., \cite{balzer_lnp13},
\begin{align}
\label{gdef}
 g_{ss'}^\sigma(t,t')&=-\ii\mean{\tc\,\op{c}_{s,\sigma}(t)\,\op{c}^\dagger_{s',\sigma}(t')}\\
 &=\thc(t-t')\,g_{ss'}^{\sigma,>}(t,t')+\thc(t'-t)\,g_{ss'}^{\sigma,<}(t,t')\;,\nn
\end{align}
with the site indices $s$, $s'$ and the spin projection $\sigma$ which attains the values $\sigma=\uparrow,\downarrow$ (we assume that there are no spin-dependent contributions to the hamiltonian). Here and below we use atomic units with $\hbar=1$. Further, the operator $\tc$ accounts for contour ordering of the times $t$ and $t'$, and $\mean{ \ldots}$ means averaging in the grand canonical ensemble.
From the NEGF all relevant observables can be computed. In particular, the density matrix follows from the time-diagonal components, $\rho_{ss'}^{\sigma}(t)=-\ii g_{ss'}^{\sigma,<}(t,t)$. In the site-diagonal case this quantity contains the expectation value of the density operator, $\rho_{ss}^{\sigma}(t)=\langle \op{n}_{s,\sigma}(t) \rangle $ introduced above whereas the off-diagonal elements are related to transition amplitudes between different sites. In a similar manner the NEGF yields probability currents, mean energies and other relevant quantities, e.g., \cite{balzer_lnp13}. 
%or equivalently $\rho_{ss'}^{\sigma,>}(t)=-\ii g_{ss'}^{\sigma,>}(t,t)$.

Due to the two-time dependence of the Green function, a systematic treatment of dynamic correlation effects is possible where interactions, quantum and spin effects as well as coupling to a (possibly strong) external field is properly taken into account. The NEGF formalism provides the basis for a theory that maintains the conservation laws \cite{book_kadanoffbaym_qsm} and allows for a systematic construction of approximations via Feynman diagrams.

Let us now consider the equations of motion for the nonequilibrium Green functions (\ref{gdef})---the Keldysh-Kadanoff-Baym equations (KBE)~\cite{book_kadanoffbaym_qsm,balzer_lnp13},
\begin{equation}
\label{kbe}
 \left(-\ii\frac{\partial}{\partial t}\delta_{s\bar{s}}-h^\sigma_{s\bar{s}}(t)\right)g_{\bar{s}s'}^\sigma(t,t')=\dc(t-t')\delta_{ss'}+\intc{\bar{t}}\Sigma^\sigma_{s\bar{s}}(t,\bar{t})g_{\bar{s}s'}^\sigma(\bar{t},t')\;,
\end{equation}
where $g$ simultaneously obeys the adjoint equation with $t\leftrightarrow t'$, and summation over the repeated site index $\bar{s}$ is implied on the left and right sides. In the present case of Hubbard clusters, the single-particle hamiltonian $h^\sigma_{ss'}$ contains the kinetic energy term of Eq.~(\ref{h0}) and the perturbation $\op{H}_1$. The two-particle hamiltonian of Eq.~(\ref{h0}) is accounted for by the one-particle self-energy $\Sigma^\sigma_{ss'}(t,t')$ appearing on the right hand side of the equation (the selfenergy contains a time-diagonal part---the Hartree-Fock selfenergy---that can be taken out of the integral and a time non-local ``correlation'' part that gives rise to the collision integral). Note that, in Eq.~(\ref{kbe}), the time arguments are defined on the Keldysh contour meaning that the functions $g$ and $\Sigma$ possess an internal matrix structure depending on how the time arguments are positioned on the contour $\cc$, for details see Ref.~\cite{balzer_lnp13}. For the following it is sufficient to note that the simulations yield, among others, the correlation functions $g_{ss'}^{\sigma,\gtrless}(t,t')$ which determine all relevant time-dependent observables, as discussed above.

The KBE are---in principle---exact equations of motion of the many-body system would the selfenergy be exactly known. This is the case only for a limited number of models. In general, therefore, one has to resort to many-body approximations for the self-energy. Due to the existence of diagram expansions, this can be done in a systematic and conserving way with the remarkable property that the approximations remain fully valid in nonequilibrium, including slow and rapid processes as well as weak and strong excitation. The simplest approximation is the Hartree-Fock (HF) approximation where correlations are neglected entirely. It is commonly expected that this is a reasonable approximation for weak coupling, i.e., in the present case, for $U\ll 1$. Nevertheless, we will see below that even for small $U$, in some nonequilibrium situations correlation effects may play a crucial role, in particular, for the long-time behavior. Among the higher order selfenergies we mention the second(-order) Born (2B), GW or T-matrix approximations \cite{book_bonitz_qkt}. In this paper we will focus on the second order Born approximation. For the treatment of Hubbard nano-clusters in higher order approximations we refer to Ref.~\cite{puigvonfriesen10}.
 
The solution of the KBE (\ref{kbe}) is now routine, e.g., \cite{book_bonitz_qkt,book_bonitz_semkat,balzer_lnp13} and references therein. After preparing a correlated initial state e.g., \cite{semkat_jmp00,stefanucci13} the system is propagated in the two-time plane by computing the NEGF as a function of both time arguments. Due to the time-memory structure of the collision integral in Eq.~(\ref{kbe}) the NEGF at all times and for all values of the site and spin indices has to be stored in memory \cite{fedvr}. Here substantial advances could be recently achieved via sophisticated program structure and parallelization~\cite{balzer_pra10,balzer_pra10_2}. Nevertheless, the computational requirements for the KBE solutions exhibit an unfavorable cubic scaling with time \cite{hermanns12_pysscripta}. Clearly, this limits the duration of propagation in nonequilibrium as well as the accuracy and resolution of the computed energy spectra that are obtained from a Fourier transform (time integral over the whole simulation). 

To overcome this limitation, we have recently developed solutions of the KBE in the single-time limit.
This is achieved by applying the generalized Kadanoff Baym ansatz (GKBA)~\cite{lipavsky86,hermanns12_pysscripta}, where the two-time functions appearing in the collision integral of Eq.~(\ref{kbe}) are ``reconstructed'' from their values on the time-diagonal according to
\begin{align}
\label{gkbadef}
 g_{ss'}^{\sigma,\gtrless}(t,t')=-g^{\sigma,\mret}_{s\bar{s}}(t,t')\,\rho_{\bar{s}s'}^{\sigma,\gtrless}(t')+\rho_{s\bar{s}}^{\sigma,\gtrless}(t)\,g^{\sigma,\madv}_{\bar{s}s'}(t,t')\;,
\end{align}
where summation over $\bar{s}$ is implied and we denoted $\rho_{\bar{s}s'}^{\sigma, <}(t)=\rho_{\bar{s}s'}^{\sigma}(t)$, and $\rho_{\bar{s}s'}^{\sigma, >}(t)=1 \pm \rho_{\bar{s}s'}^{\sigma}(t)$,  where ``+'' (``-'') refers to bosons (fermions).
For an explicit expression for $\Sigma^\sigma_{ss'}(t,t')$ as a functional of the NEGF and for the corresponding collision term, see, e.g., Ref.~\cite{hermanns_jpcs13}. Finally, the two-time retarded and advanced propagators $g^{\sigma,\mret}_{ss'}(t,t')$ and $g^{\sigma,\madv}_{ss'}(t,t')$ are computed in Hartree-Fock approximation rendering the ansatz highly efficient,
\begin{align}\label{eq:hf_prop}
 g^{\sigma,\mret/\madv}_{ss'}(t,t')=\left.\mp\ii\thc(\pm[t-t'])\exp{\left(-\ii\intlim{t'}{t}{\bar{t}}h^\sigma_\mathrm{HF}(\bar{t})\right)}\right|_{ss'}\;.
\end{align}
Here, $h^\sigma_\mathrm{HF}(t)$ denotes the single-particle time-dependent Hartree-Fock hamiltonian.
We emphasize that the reconstruction of the greater and lesser components of the Green function with real $t$ and $t'$ is sufficient as long as the method of adiabatic switching is applied to generate the correlated initial (ground) state by time propagation, for details see \cite{hermanns12_pysscripta}. The quality of the GKBA has been tested before for macroscopic spatially homogeneous systems \cite{bonitz_jpcm96}. There it was found that the GKBA retains the conservation laws of the original two-time approximation for the selfenergy \cite{book_bonitz_qkt, bonitz_pla96}. Furthermore, it was found that this ansatz is a very good approximation to the full two-time solution if the exact propagators $g^{\mret/\madv}(t,t')$ are being used and the results remain satisfactory with the Hartree-Fock propagators \cite{kwong98}. The use of damped propagators that include imaginary selfenergy contributions, on the other hand, violates total energy conservation and leads to an overall worse performance \cite{bonitz_epjb99}.
Our recent results for lattice systems confirm these observations and indicate that the second Born-GKBA (2B-GKBA) with HF propagators provides an excellent description of Hubbard nano-clusters up to moderate couplings of the order $U\sim 1$. We underline that the scaling with the simulation time $T_{\textnormal{s}}$ was found to improve to $\mathcal{O}\left(T_{\textnormal{s}}^2\right)$ \cite{hermanns_jpcs13}---a noticeable gain for the desired long relaxation studies compared to full two-time simulations.

%In general, the application of MBAs and other simplifications such as the GKBA require thorough justification as they can lead to unphysical effects such as self-interaction errors and spurious dynamical excitations~\cite{hermanns12_prb}, bistability~\cite{khosravi12} or artificial steady states~\cite{puigvonfriesen10}. For this reason, we, in this contribution, extend previous work on weak perturbations~\cite{balzer12_epl}, which included the discussion of double excitations (see also~\cite{sakkinen12}), to the nonlinear regime and analyze the performance of the GKBA for a finite system with the two-body interactions being treated in the second Born approximation.

\section{Numerical Results}\label{s:results}
We now apply our NEGF results within the 2B-GKBA approximation to the nonlinear excitation of a small Hubbard cluster. 
In the following, we consider the case of a strong nonlinear perturbation with $\epsilon_0=5.0$ and resort to the zero-temperature limit $\beta\rightarrow\infty$. Initially, at $t=0$, the system is prepared in the ground state of $\op{H}_0$.
Following Ref.~\cite{balzer_jpcs13}, the response of the Hubbard chain with respect to the time-dependent electron density on the first site ($s=0$) is quantified by computing
\begin{align}
\label{gammadef}
 \gamma^\sigma(t)=\mean{\op{n}_0^\sigma}(t)-\frac{1}{t^*}\intlim{0}{t^*}{\bar{t}}\mean{\op{n}_0^\sigma(\bar{t})}\;,\\
 \label{gammadefft}
  \gamma^\sigma(\omega)=\intlim{0}{t^*}{t}\,\gamma^\sigma(t)\,e^{-\ii\omega t}\;,
\end{align}
where $\mean{\op{n}_0^\sigma}(t)=-\ii g_{00}^{\sigma,<}(t,t)$. With the definition (\ref{gammadef}) we remove a (possibly large) average contribution to the density from the time-dependent observable $\mean{\op{n}_0^\sigma}(t)$ (resulting in a large zero-frequency peak in the spectrum, see, e.g., Fig.~\ref{fig3}).
The time $t^*$ indicates a finite propagation time used in the numerics and is chosen sufficiently large such that it only affects the basic width of the peaks in $\gamma(\omega)$ but not their position. Furthermore, due to the spin symmetry obeyed by \eq{h1}, we have $\gamma(\omega)=\gamma^\uparrow(\omega)=\gamma^\downarrow(\omega)$. 
\begin{figure}[t]
 \includegraphics[width=0.89\textwidth]{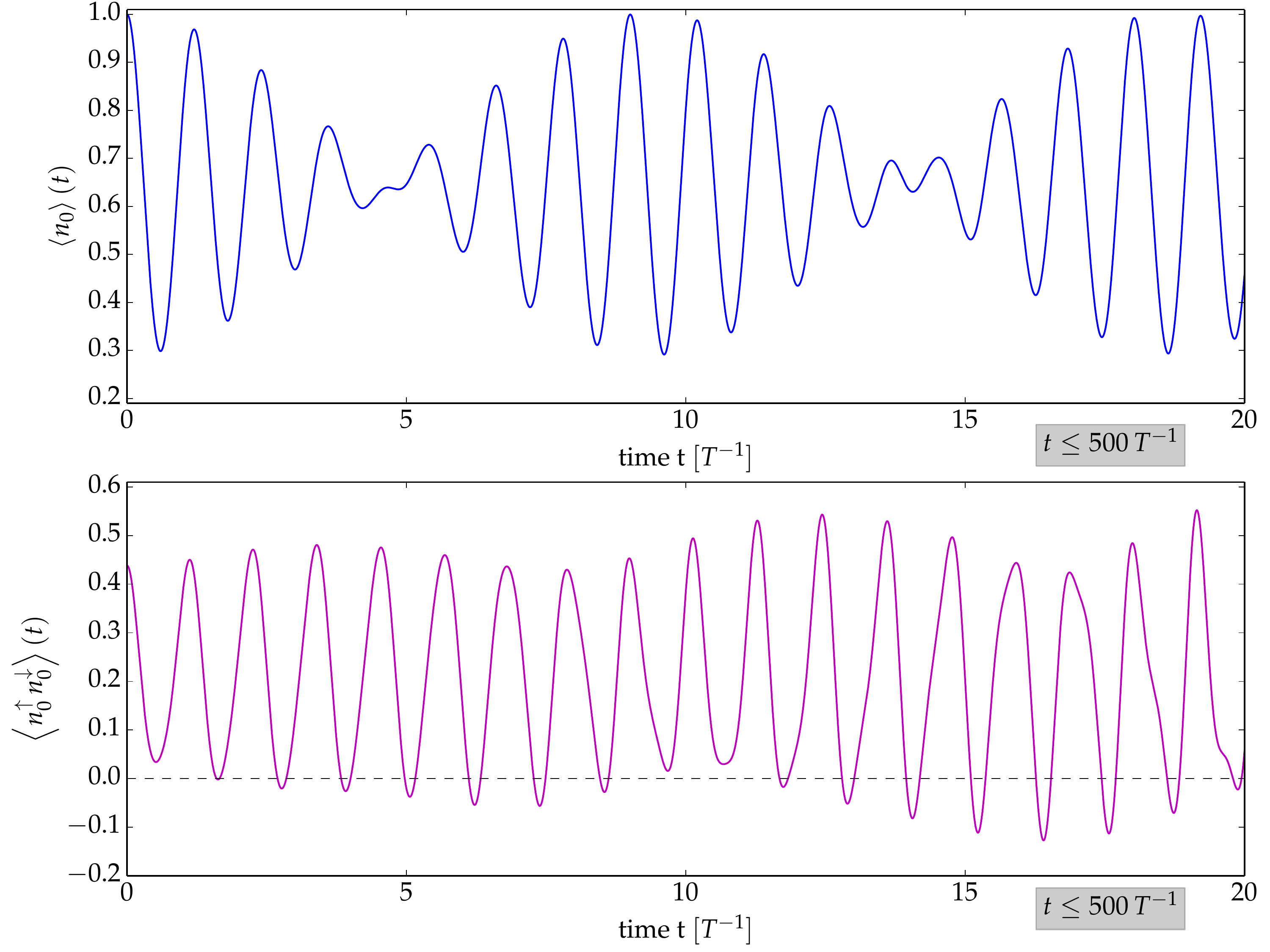}
 \caption{Time-dependent density on site ``0'' caused by the excitation of the 2-site Hubbard cluster with the hamiltonian $\op {H}_1$, Eq.~(\ref{h1}) for a strength of $\epsilon_0=5.0$, at half filling ($N=2$) for $U=0.5$. The lower figure shows the time evolution of the double occupancy on site ``0''. The whole simulation has a duration of $500\,T^{-1}$.}
 \label{fig2}
\end{figure}

In Fig.~\ref{fig2} we show the time dependence of the occupation $\left< n_0\right>$ and the double occupation $\left< n_0^\uparrow n_0^\downarrow \right>$ of site ``0'' computed from the solution of the KBE using the second Born approximation and the GKBA (for a comparison with the results from exact diagonalization, we refer to Ref.~\cite{balzer_jpcs13}). For both quantities, one observes a periodic time dependence that is characterized by several frequencies. With respect to the spectrum, the double occupancy shows to be more mono-chromatic compared to the single occupancy. Additionally, one notices that the double occupancy partially assumes slightly negative values (crossing the dashed line) which is unphysical but does not seem to influence the stability of the propagation algorithm negatively. For a recent discussion of this issue we refer to Ref.~\cite{friesen_10}. To better understand the dynamics, we show the Fourier-transformed results in Figure~\ref{fig3}. The dashed (thick) curves depict the response $\gamma(\omega)$ (Gaussian fits of it) for a two-site chain ($L=2$) at different repulsive interaction strengths $U$ according to \eq{gammadefft}.  The quasi-non-interacting system ($U=0.01$, black curve)  exhibits two peaks---one at $\omega_0=5.385$ and one at $2\omega_0=10.770$. Interestingly, for $U>0$,  the energetically lowest peak at $\omega_0$ splits into two separated peak structures where the right steadily gets a smaller spectral weight. With further increase of $U$ the difference of the spectral weights of the two structures vanishes, cf.~Figure~\ref{fig3} for $U=1.5$. In contrast, the peak at $2\,\omega_0$ changes only weakly with $U$ with an monotonically increasing spectral weight for $U \le 1.5$. The transition energies are plotted again in Fig.~\ref{fig4} over a broader range of interaction strengths $U$. There, the positions of all transitions have been averaged by Gaussians over all contributing peaks, see Fig.~\ref{fig3}.
\begin{figure}[t]
 \includegraphics[width=0.89\textwidth]{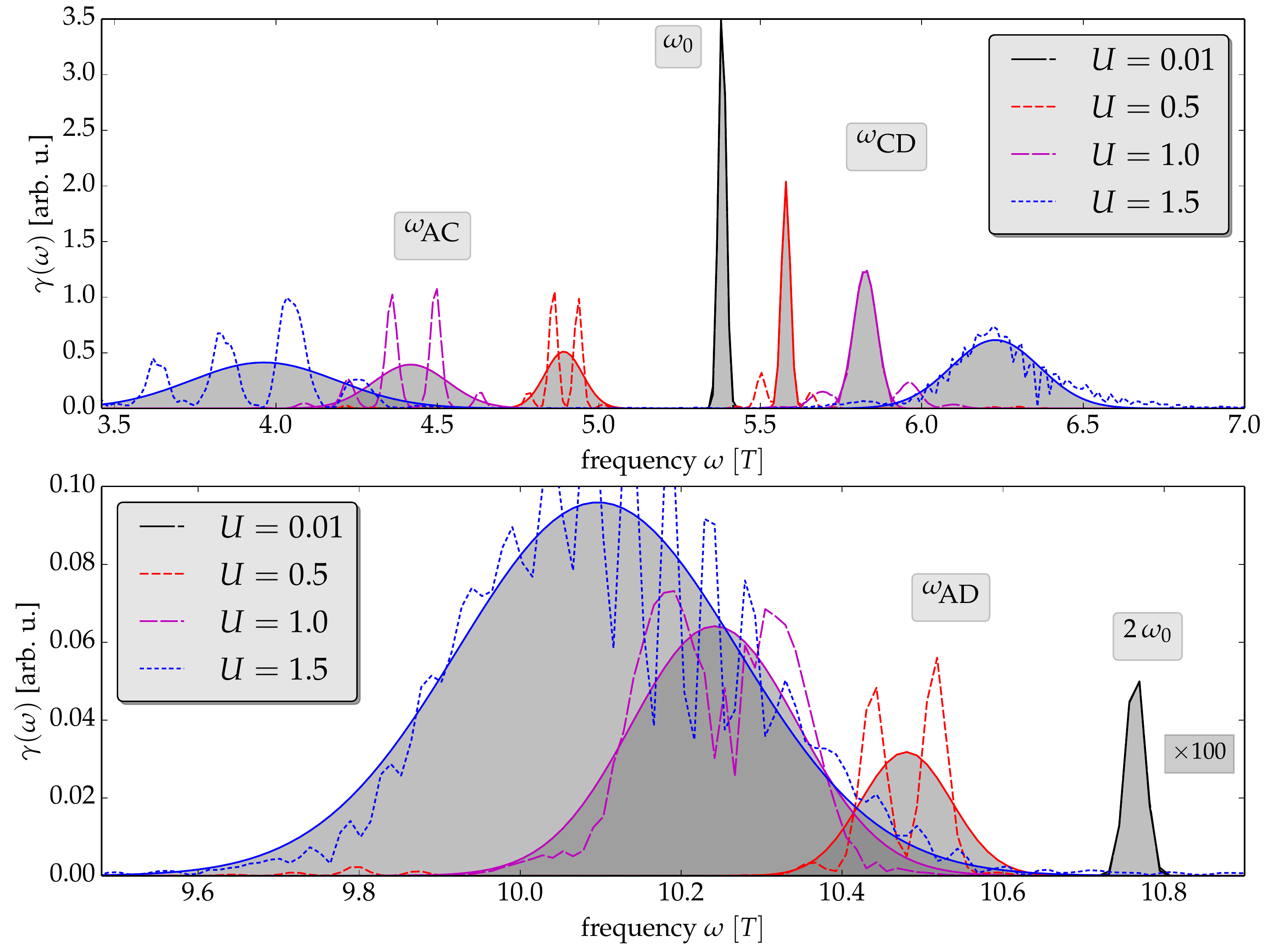}
 \caption{Nonlinear density response spectra , \eq{gammadefft}, of site ``0'', $\gamma(\omega)=\gamma^\uparrow(\omega)=\gamma^\downarrow(\omega)$, for a two-site chain at $\epsilon_0=5.0$ and half-filling computed from the NEGF within 2B-GKBA. The dotted lines represent the calculated data while the shaded areas under the full lines are Gaussian fits. The lower (upper) part of the resulting spectra is plotted in the upper (lower) figure for four representative interaction strengths $U$. In the lower figure, the peak for $U=0.01$ has been amplified by a factor of $100$ for better visibility.}
 \label{fig3}
\end{figure}

To understand the spectrum \cite{balzer_jpcs13}, we recall that we are studying a switch between two time-independent hamiltonians given by \eqsand{h0}{h1} that occurs instantaneously at time $t=0$, without any ramp function. In general, the initial state (the ground state of $\op{H}_0$) will not be an eigenstate of the new hamiltonian, $\op{H}_{2}=\op{H}_0+\op{H}_1$, for $t>0$, but can be expressed as a superposition of the eigenstates of the latter. Consequently, the nonequilibrium dynamics of the system are governed by the transition frequencies between the eigenstates of $\op{H}_{2}$ which should show up in the spectrum $\gamma(\omega)$. 
\begin{figure}[t]
 \includegraphics[width=0.89\textwidth]{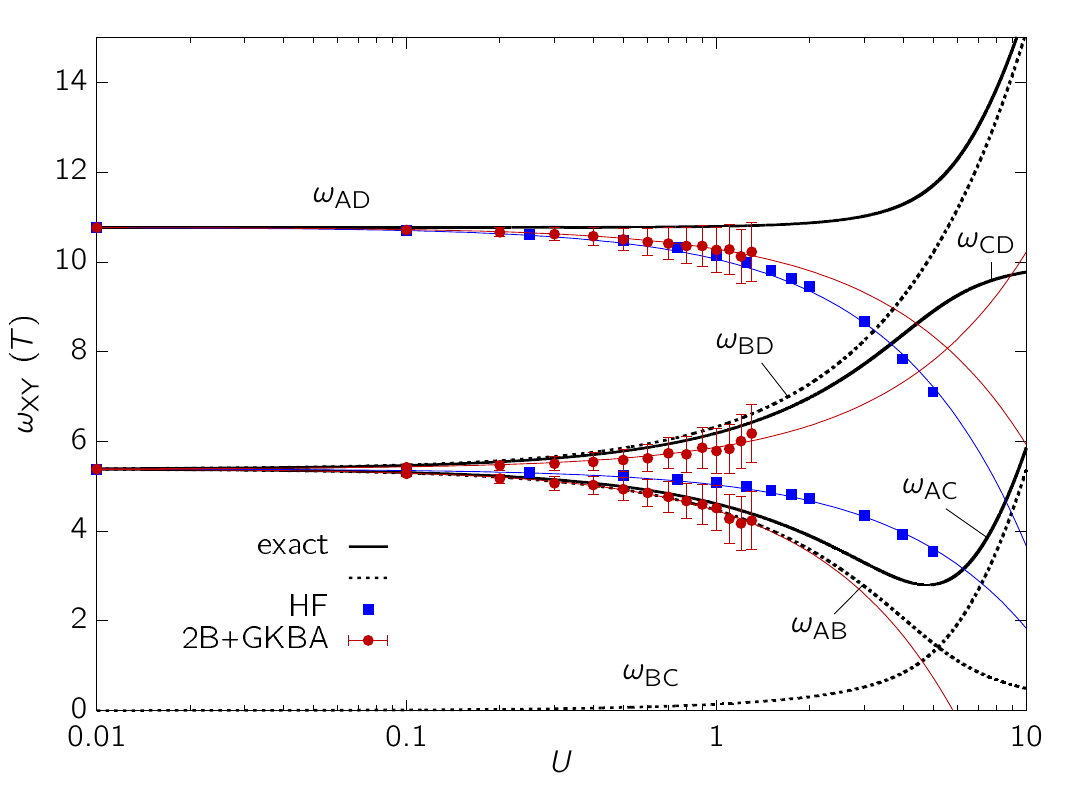}\hspace{-0.5pc}
 \caption{Comparison of the exact transition frequencies $\omega_{\mathrm{AC}}$, $\omega_{\mathrm{CD}}$ and $\omega_{\mathrm{AD}}$ (black solid lines, Ref.~\cite{balzer_jpcs13}) to the results obtained in HF (blue, squares) and 2B+GKBA (red, dots). The thin (blue and red) lines are linear fits to the data points according to the formulas given in the text. The error bars indicate the width of the peaks and arise from the finite simulation duration (finite time interval in the Fourier transform).}
 \label{fig4}
\end{figure}

Figure~\ref{fig1}~b) shows the eigenenergies in the asymmetric chain with $\epsilon_0=5.0$ as a function of $U$, and the thin vertical lines indicate the associated excitation frequencies which can be exited from the groundstate with $\epsilon_0=0$, which is explained in the following. From an analysis of the exact eigenstates of the two hamiltonians $\op{H}_0$ and $\op{H}_2$ it is known that there is finite overlap only between the ground state of $\op{H}_0$ and the states $\mathrm{A}$, $\mathrm{C}$ and $\mathrm{D}$ of $\op{H}_2$ (all being singlets), cf. the sketch in Fig.~\ref{fig1}~a). The vanishing overlap with the eigenstate $\mathrm{B}$ (the triplet with $S_z=0$,~\cite{jafari08}) is responsible for the fact that we do not observe transitions involving state $\mathrm{B}$ in the density response. In terms of the wave function, the dynamics of the system for $t>0$ is therefore,
\begin{align}
 \ket{\Psi(t)}=c_\mathrm{A}\,\e{-\ii E_\mathrm{A}t}\ket{\mathrm{A}}+c_\mathrm{C}\,\e{-\ii E_\mathrm{C}t}\ket{\mathrm{C}}+c_D\,\e{-\ii E_\mathrm{D}t}\ket{\mathrm{D}}\;,
\end{align}
with $\ket{\mathrm{X}}$ being the eigenstates of $\op{H}_2$ (having energy $E_\mathrm{X}$) and $c_\mathrm{X}=\braket{\mathrm{X}}{\mathrm{gs}}$ denoting the expansion coefficients with respect to the ground state $\ket{\mathrm{gs}}$ of $\op{H}_0$. 
Moreover, it is easily shown~\cite{balzer_jpcs13} that the excited state $\mathrm{D}$ is a doubly-excited state relative to the ground state $\mathrm{A}$, whereas the states $\mathrm{B}$ and $\mathrm{C}$ are singly-excited states. With the same argument, state $\mathrm{D}$ is furthermore also a doubly-excited state relative to the ground state (gs) of $\op{H}_0$. We note that, in the NEGF formalism we do not have direct access to the $N$-particle states, nevertheless the characteristic frequencies are captured by the spectra (spectral function, propagators) or the associated time dynamics of the relevant observables such as the site occupation. Thus it is obvious that the peaks in the Fourier transform of the density response, cf. Fig~\ref{fig3}, should within the approximation coincide with the transitions in $\op{H}_1$. Indeed, we find the approximations of the transitions $\omega_\mathrm{AC}$, $\omega_\mathrm{CD}$ in the upper part of Fig.~\ref{fig3} and of $\omega_\mathrm{AD}$ in the lower part.

Let us now analyze the role of correlation effects. This can be easily done by turning off correlations in the NEGF scheme entirely by neglecting the collision integral, i.e., by retaining in the selfenergy only the time-diagonal Hartree-Fock contribution. The corresponding results are shown in Fig.~\ref{fig4} by the (blue) squares. The analysis shows that the transition with the highest energy difference $\mathrm{A}\leftrightarrow\mathrm{D}$ is missing in the HF solution because double excitations are generally not included in any time-dependent Hartree-Fock calculation~\cite{balzer12_epl}.
 In contrast, the two transitions $\mathrm{A}\leftrightarrow\mathrm{C}$ and $\mathrm{C}\leftrightarrow\mathrm{D}$ are of one-electron character, however, only the transition $\mathrm{A}\leftrightarrow\mathrm{C}$ is observed. The reason is that the function $\gamma(\omega)$---combined with the excitation (\ref{h1})---only probes energy differences between states that are populated already at time $t=0$. As mentioned above, however, the state $\mathrm{D}$ is never populated in HF. For this reason, the splitting of the low energy peak around $\omega=5.4$ observed in the 2B-GKBA simulations, cf. Fig.~\ref{fig3}, is completely missing in a HF simulation. Inspection of the HF curves in Fig.~\ref{fig4} shows, besides the curve corresponding to the transition $\mathrm{A}\leftrightarrow\mathrm{C}$, a second line close to the second Born transition $\mathrm{A}\leftrightarrow\mathrm{D}$. This appears to be in conflict with the analysis of double excitations given above. However, there is a simple explanation: by performing a linear fit to the two HF frequencies we obtain $\omega_\mathrm{AC}^\mathrm{HF}=-0.355\,U+5.385\;$ and
 $\omega_\mathrm{AD}^\mathrm{HF}=-0.710\,U+10.770=2\,\omega_\mathrm{AC}^\mathrm{HF}\;$. This means, the observed frequency around $\omega_\mathrm{AD}$ is just the second harmonics of the frequency $\omega_\mathrm{AC}$.

In contrast to the HF approximation, NEGF-simulations with second-order self-energy are able to reproduce double excitations, as was shown in Refs.~\cite{balzer12_epl,sakkinen12}. Therefore, the present 2B+GKBA calculations capture the transition $\omega_\mathrm{AD}$ and, accordingly, also $\omega_\mathrm{CD}$. The fact that the transition $\mathrm{A}\leftrightarrow\mathrm{D}$ is not the second harmonic of $\mathrm{A}\leftrightarrow\mathrm{C}$, as in the HF case, is readily verified by making an analogous linear fit through the simulation points 
with the result \cite{balzer_jpcs13}
$  \omega_\mathrm{AC}^\mathrm{2B}=-0.928\,U+5.385\;,\hspace{1pc}
    \omega_\mathrm{CD}^\mathrm{2B}=0.483\,U+5.385\;,\hspace{1pc}
      \omega_\mathrm{AD}^\mathrm{2B}=-0.385\,U+10.770\;$.
%This gives, for the sum, $\omega_\mathrm{AC}^\mathrm{2B}+\omega_\mathrm{CD}^\mathrm{2B} \approx -0.445 \, U +10.7$ which differs significantly from $\omega_\mathrm{AD}^\mathrm{2B}$.
The difference between $\omega_\mathrm{AD}^\mathrm{2B}$ and the second harmonic of $\omega_\mathrm{AC}^\mathrm{2B}$ grows with $U$ as 
$\omega_\mathrm{AD}^\mathrm{2B}-2 \omega_\mathrm{AC}^\mathrm{2B} \approx 1.27 U$.

Finally, we note that our 2B+GKBA simulations reveal also another feature of small Hubbard clusters---the possibility of finite double occupations of a given site. The corresponding time-dependent results are included in the lower part of Fig.~\ref{fig2}. The dynamics of the double occupation $\langle {\hat n}_i^\uparrow {\hat n}_i^\downarrow \rangle(t)$ was computed according to the formula \cite{puigvonfriesen10} $\langle {\hat n}_i^\uparrow {\hat n}_i^\downarrow \rangle(t)=-\textnormal{i}U^{-1}\left[\int_{\mathcal{C}}\textnormal{d}3\Sigma(13) G(31^+)\right]_{ii}$, which is essentially the diagonal element of the two-particle correlation function. We observe a similar oscillatory time-dependence as for the single occupations (upper part of the figure). The carrier frequency (highest frequency) is the same in both quantities. However, the oscillations of the double occupations are much less modulated, except for the already mentioned slight violation of positivity.
Apart from this, the results are in good agreement with the exact data.

Thus, our second order Born (2B+GKBA) results give a correct picture of the main features of the spectrum of the two-site Hubbard cluster, even in the case of strong nonlinear excitation. Obviously, the treatment of correlations on this level is only an approximation, and we should expect increasing deviations for growing $U$. This can be clearly seen in Fig.~\ref{fig5} where we included also results from exact diagonalization (full black lines).
While the two lower frequencies $\omega_\mathrm{AC}$ and $\omega_\mathrm{CD}$ are very well reproduced by 2B+GKBA, the upper mode $\omega_\mathrm{AD}$ is only rather accurate for $U \lesssim 1.5$. Additionally, it is obvious that it exhibits an incorrect slope with $U$: while the exact result shows an increase of the frequency with $U$ the Born approximation yields a decrease \cite{balzer_jpcs13}. To reproduce the correct behavior, obviously, higher order correlation contributions are essential.
\section{Conclusion and Outlook}
This paper was devoted to Hubbard nano-clusters---finite lattice systems of electrons that are well suited to study correlation effects in quantum systems in combination with out-of-equilibrium behavior following an external excitation. Here we concentrated on the dynamics of the system triggered by imposing a strong external potential at time $t=0$, which results in a pronounced nonequilibrium particle distribution across the sites. A similar scenario, where both electrons were artificially placed on the same site initially, was studied before using NEGF \cite{puigvonfriesen10} in order to test the method and different approximations. The authors of that reference observed unphysical relaxation behavior---a strong damping in the system that is not present in the exact solution. Our approach that uses, in addition to the second Born approximation, the GKBA does not exhibit these problems which is quite encouraging. A similar observation of the suitability of the second Born approximation with the GKBA was reported in Ref~\cite{hermanns_jpcs13} where our results behave favorably over a very long time whereas density matrix results of Akbari \textit{et al.} \cite{akbari_12} experienced instabilities and other serious problems. The origin of this improved behavior is the use of undamped HF propagators, c.f. Eq.~(\ref{eq:hf_prop}), which does not alter the (second) order of the correlation effects but improves the width of the peaks in the spectrum which is essential for finite systems.
\begin{figure}[t]
 \includegraphics[width=0.99\textwidth]{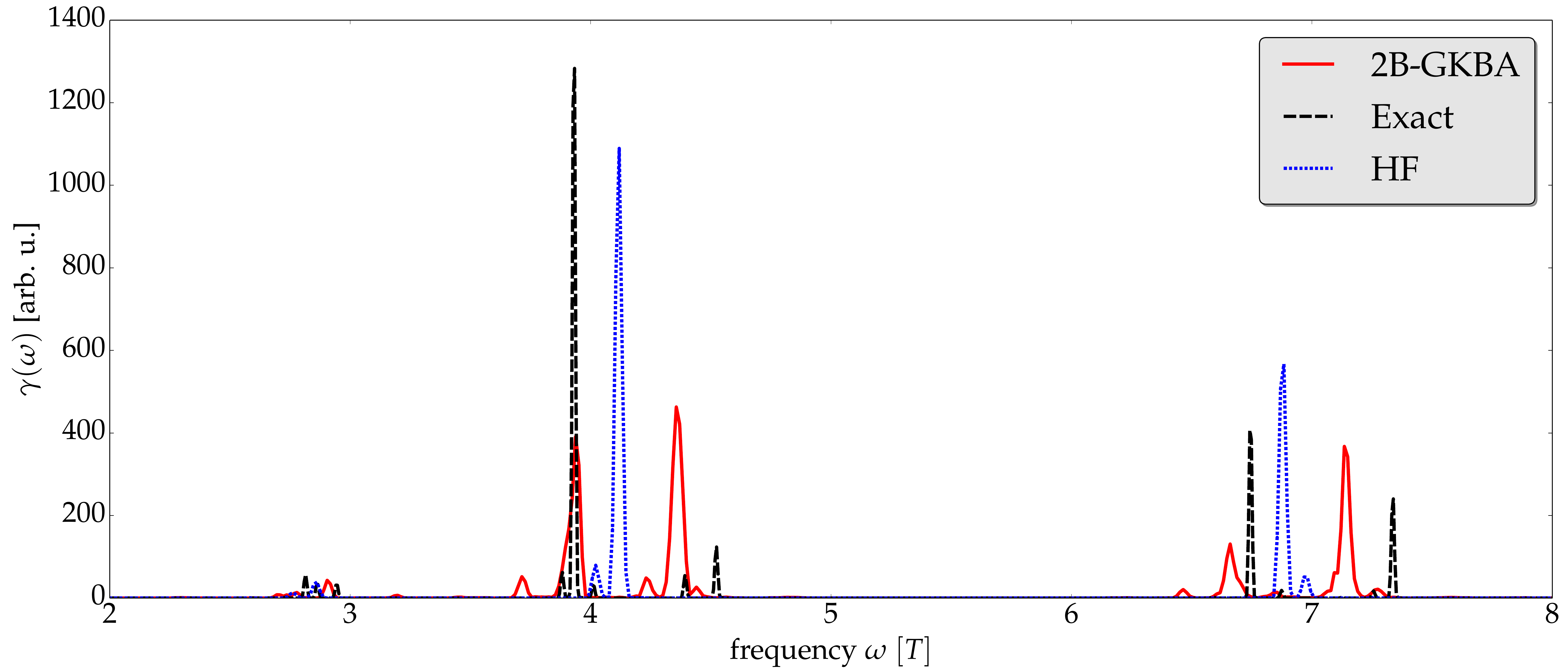}\hspace{-0.5pc}
 \caption{Nonlinear density response spectrum (low-energy part) of site ``0'', $\gamma(\omega)$, for a 2D Hubbard cluster with $2\times 2$ sites at half filling ($N=4$) and $U=0.5$ for the same excitation conditions as in Fig.~\ref{fig1}. Besides the 2B-GKBA data we show results from exact diagonalization and from HF-simulations. Note that for the first peak, the exact and the 2B-GKBA results lie exactly on top of each other and that in the HF case there is no splitting of the two upper peaks.}
 \label{fig5}
\end{figure}

Besides the favorable physical behavior of our approximation in the time evolution of Hubbard nano-clusters we note that our NEGF approach has a very attractive scaling with the number of particles which will allow us to approach substantially larger systems in the near future \cite{hermanns_tobepublished}. As a first illustration we show the density response of a 2D cluster with $2\times 2$ sites at half filling ($N=4$) for the same excitation conditions as used above. For this (still small) system we have exact diagonalization results available that allow to gauge the quality of our results. The results are shown in Fig.~\ref{fig5}. The low-energy part of the spectrum contains three peaks where the lowest one has only a low occupation. When $U$ is increased the two upper peaks both split in two. This splitting is not observed for $U=0$, and it is missing in Hartree-Fock for all $U$, i.e., it is a correlation effect. In contrast, our 2B-GKBA results correctly reveal the splitting of both peaks and also show the correct trend of increasing splitting when $U$ is increased (not shown). We generally observe that the lower (higher) peak of each doublet is reproduced very (slightly less) accurately (compare to the exact data in Fig.~\ref{fig5}).

We performed an additional series of simulations for larger systems that fully confirmed the feasibility of larger clusters. We studied the required CPU time to simulate from $N=8$ to $N=32$ sites. Exact diagonalization results were only available for $N\le 8$. This is caused by the exponential scaling of this method, i.e., the CPU time scales as $\mathcal{O}\left(\left(N_\textnormal{b}\right)^N\cdot T_{\textnormal{s}}\right)$, where $N_\textnormal{b}$ is the basis dimension. 
Alternatively, we applied multiconfiguration time-dependent Hartree-Fock simulations (MCTDHF) \cite{mctdhf,hochstuhl_jcp11} that scale like $\mathcal{O}\left(M^N\cdot T_{\textnormal{s}}\right)$, where $M$ is the number of time-dependent orbitals, making the scaling independent of the basis size but still suffering from the exponential scaling with $N$. In striking contrast, our 2B-GKBA nonequilibrium Green functions simulations scale as $\mathcal{O}\left(N_\textnormal{b}^4\cdot T_{\textnormal{s}}^2\right)$, completely independent of the number of particles. This gives us confidence that substantially larger systems can be treated by our method \cite{hermanns_tobepublished}.

\begin{acknowledgement}
 This work was supported by the Deutsche Forschung Gemeinschaft via grant BO1366-9 and the Northern German Supercomputing Alliance (HLRN) via grant shp006.
\end{acknowledgement}

%----------------------


\begin{thebibliography}{10}

\bibitem{pavarini11}{E.~Pavarini, E.~Koch, D.~Vollhardt and A.~Lichtenstein (Eds.), \textit{The LDA+DMFT approach to strongly correlated materials}, (Forschungszentrum J\"ulich GmbH, Zentralbibliothek, Verlag 2011).}

\bibitem{balzer_jpcs13} K. Balzer, S. Hermanns, and M. Bonitz, J. Phys. Conf. Ser. {\bf 427}, 012006 (2013)

\bibitem{friesen_10}{M.~Puig von Friesen, C.~Verdozzi and C.-O.~Almbladh,
\textit{Kadanoff-Baym equations and approximate double occupancy in a Hubbard dimer}, 
arXiv:1009.2917 (2010).}

\bibitem{becker12}{W.~Becker, X.J.~Liu, P.J.~Ho and H.J.~Eberly, 
%\textit{Theories of photoelectron correlation in laser-driven multiple atomic ionization}, 
Rev.~Mod.~Phys.~\textbf{84}, 1011 (2012).}

\bibitem{schuette_prl12} B. Sch\"utte, S. Bauch, U. Fr\"uhling, M. Wieland, M. Gensch, E. Pl\"onjes, T. Gaumnitz, A. Azima, M. Bonitz, and M. Drescher,
%Evidence for chirped Auger electron emission
Phys. Rev. Lett. {\bf 108}, 253003 (2012); S. Bauch, and M. Bonitz, Phys. Rev. A {\bf 85}, 053416 (2012)

\bibitem{bloch08}{I.~Bloch, J.~Dalibard and W.~Zwerger, 
%\textit{Many-body physics with ultracold gases}, 
Rev.~Mod.~Phys.~\textbf{80}, 885 (2008).}

\bibitem{wall11}{S.~Wall, D.~Brida, S.R.~Clark, H.P.~Ehrke, D.~Jaksch, A.~Ardavan, S.~Bonora, H.~Uemura, Y.~Takahashi, T.~Hasegawa, H.~Okamoto, G.~Cerullo and A.~Cavalleri, 
%\textit{Quantum interference between charge excitation paths in a solid state Mott insulator}, 
Nature Physics \textbf{7}, 114 (2011).}

\bibitem{book_bonitz_qkt}{M.~Bonitz: \textit{Quantum Kinetic Theory} (Teubner, Stuttgart/Leipzig, 1998).}

\bibitem{akbari_12} A. Akbari, M.J. Hashemi, A. Rubio, R.M. Nieminen, and R. van Leeuwen, Phys. Rev. B {\bf 85}, 235121 (2012)

\bibitem{hermanns_jpcs13} S. Hermanns, K. Balzer, and M. Bonitz,
%Few-particle quantum dynamics - comparing Nonequilibrium Green functions with the generalized Kadanoff-Baym ansatz to density operator theory
J. Phys. Conf. Ser. {\bf 427}, 012008 (2013)

\bibitem{bonitz_cpp99} M. Bonitz, Th. Bornath, D. Kremp, M. Schlanges, and W.D. Kraeft,
%Quantum kinetic theory for laser plasmas. Dynamical screening in strong fields,
Contrib. Plasma Phys. {\bf 39}, 329 (1999)

\bibitem{haberland_pre01} H. Haberland, M. Bonitz, and D. Kremp,
%Harmonics generation in electron-ion collisions in a strong laser pulse
Phys. Rev. E {\bf 64}, 026405 (2001)

\bibitem{kwong98}{N.H.~Kwong, M.~Bonitz, R.~Binder and H.S.~K\"ohler, 
%\textit{Semiconductor Kadanoff-Baym equation results for optically excited electron-hole plasmas in quantum wells}, 
phys.~stat.~sol.~(b)~\textbf{206}, 197 (1998).}

\bibitem{kwong00}{N.-H.~Kwong and M.~Bonitz, 
%\textit{Real-time Kadanoff-Baym approach to plasma oscillations in a correlated electron gas}, 
Phys.~Rev.~Lett.~\textbf{84}, 1768 (2000).}

\bibitem{rios11}{A.~Rios, B.~Barker, M.~Buchler and P.~Danielewicz, 
%\textit{Towards a nonequilibrium Green function description of nuclear reactions: One-dimensional mean-field dynamics}, 
Annals of Physics~\textbf{326}, 1274 (2011).}

\bibitem{garny11}{M.~Garny, A.~Kartavtsev and A.~Hohenegger, 
%\textit{Leptogenesis from first principles in the resonant regime}, arXiv:1112.6428.
Annals of Physics {\bf 328}, 26 (2013)}

\bibitem{gartner06}{P.~Gartner, J.~Seebeck and F.~Jahnke, 
%\textit{Relaxation properties of the quantum kinetics of carrier-LO-phonon interaction in quantum wells and quantum dots}, 
Phys.~Rev.~B~\textbf{73}, 115307 (2006).}

\bibitem{lorke06}{M.~Lorke, T.R.~Nielsen, J.~Seebeck, P.~Gartner and F.~Jahnke, 
%\textit{Influence of carrier-carrier and carrier-phonon correlations on optical absorption and gain in quantum-dot systems}, 
Phys.~Rev.~B~\textbf{73}, 085324 (2006).}

\bibitem{bonitz_prb07} M. Bonitz, K. Balzer, and R. van Leeuwen,
%Invariance of the Kohn (sloshing) mode in a conserving theory,
%ArXiv: cond-mat/0702633
Phys. Rev. B {\bf 76}, 045341 (2007). 

\bibitem{balzer_prb09} K. Balzer, M. Bonitz, R. van Leeuwen, N.E. Dahlen, and A. Stan,
%Nonequilibrium Green functions approach to strongly correlated few-electron quantum dots
Phys. Rev. B {\bf 79},  245306 (2009).

\bibitem{uimonen11} A.-M.~Uimonen, E.~Khosravi, A.~Stan, G.~Stefanucci, S.~Kurth and R.~van Leeuwen and E.K.U.~Gross, 
%\textit{Comparative study of many-body perturbation theory and time-dependent density functional theory in the out-of-equilibrium Anderson model},
Phys.~Rev.~B~\textbf{84}, 115103 (2011).

\bibitem{khosravi12}{E.~Khosravi, A.-M.~Uimonen, A.~Stan, G.~Stefanucci, S.~Kurth, R.~van Leeuwen and E.K.U.~Gross, 
%\textit{Correlation effects in bistability at the nanoscale: Steady state and beyond}, 
Phys.~Rev.~B~\textbf{85}, 075103 (2012).}

\bibitem{dahlen07} N.E.~Dahlen, and R. van Leeuwen, Phys. Rev. Lett. {\bf 98}, 153004 (2007). 

\bibitem{balzer_pra10} K. Balzer, S. Bauch, and M. Bonitz
%Efficient grid-based method in nonequilibrium Green’s function calculations: Application to model atoms and molecules
Phys. Rev. A {\bf 81}, 022510 (2010). 

\bibitem{balzer_pra10_2} K. Balzer, S. Bauch, and M. Bonitz
%Time-dependent second-order Born calculations for model atoms and molecules in strong laser fields
Phys. Rev. A {\bf 82}, 033427 (2010). 

\bibitem{balzer_lnp13} K. Balzer, and M. Bonitz, {\em Nonequilibrium Green Functions Approach to Inhomogeneous Systems}, 
Lecture Notes in Physics, vol. 867, Springer (2013).

\bibitem{puigvonfriesen09}{M.~Puig von Friesen, C.~Verdozzi and C.-O.~Almbladh,
%\textit{Successes and failures of Kadanoff-Baym dynamics in Hubbard nanoclusters}, 
Phys.~Rev.~Lett.~\textbf{103}, 176404 (2009).}

\bibitem{puigvonfriesen10}{M.~Puig von Friesen, C.~Verdozzi and C.-O.~Almbladh: 
%\textit{Kadanoff-Baym dynamics of Hubbard clusters: Performance of many-body schemes, correlation-induced damping and multiple steady and quasi-steady states}, 
Phys.~Rev.~B~\textbf{82}, 155108 (2010).}

\bibitem{balzer12_epl}{K.~Balzer, S.~Hermanns and M.~Bonitz,
%\textit{Electronic double-excitations in quantum wells: Solving the two-time Kadanoff-Baym equations}, 
EPL~\textbf{98}, 67002 (2012).}

\bibitem{sakkinen12}{N.~S\"akkinen, M.~Manninen and R.~van Leeuwen,
%\textit{The Kadanoff-Baym approach to double excitations in finite systems}, 
New J.~Phys.~\textbf{14}, 013032 (2012).}

\bibitem{lipavsky86} P.~Lipavsk{\'y}, V.~{\v{S}}pi{\v{c}}ka and B.~Velick{\'y}, 
%\textit{Generalized Kadanoff-Baym ansatz for deriving quantum transport equations}, 
Phys.~Rev.~B~\textbf{34}, 6933 (1986).

\bibitem{hermanns12_pysscripta} S.~Hermanns, K.~Balzer and M.~Bonitz, 
%\textit{Nonequilibrium Green functions approach to inhomogeneous quantum many-body systems using the generalized Kadanoff-Baym ansatz}, 
Physica~Scripta {\bf T151}, 014035 (2012).

\bibitem{book_bonitz_semkat} \textit{Introduction to Computational Methods in Many-Body Physics},
M. Bonitz, and D. Semkat (Eds.), Rinton Press, Princeton, 2006. 

\bibitem{semkat_jmp00} D. Semkat, D. Kremp, and M. Bonitz, 
%\textit {Kadanoff-Baym equations and non-Markovian Boltzmann equation in generalized T-matrix approximation}
J. Math. Phys. {\bf 41}, 7458 (2000).

\bibitem{stefanucci13} R van Leeuwen and G Stefanucci,
%\textit{Equilibrium and nonequilibrium many-body perturbation theory: a unified framework based on the Martin-Schwinger hierarchy} 
J. Phys. Conf. Ser. {\bf 427}, 012001 (2013).

\bibitem{fedvr} For continuous systems substantial advances have recently been achieved via the choice of special basis representations (FEDVR basis), e.g., \cite{balzer_pra10}, for lattice systems this problem does not occur.

\bibitem{bonitz_jpcm96} M. Bonitz, D. Kremp, D.C. Scott, R. Binder, W. D. Kraeft, and H. S. K\"ohler,
%Numerical analysis of memory effects in the intraband relaxation in semiconductors
Journal of Physics: Condensed Matter {\bf 8}, 6057  (1996)

\bibitem{bonitz_pla96} M. Bonitz, D. Kremp,
%Kinetic energy relaxation and correlation time of nonequilibrium many-particle systems
Phys. Lett. A {\bf 212}, 83 (1996) 

\bibitem{bonitz_epjb99} M. Bonitz, D. Semkat and H. Haug,
%Non-Lorentzian spectral functions for Coulomb quantum kinetics,
Europ. Phys. J. B {\bf 9}, 309 (1999) 

\bibitem{book_kadanoffbaym_qsm}{L.P.~Kadanoff and G.~Baym: \textit{Quantum Statistical Mechanics} (Benjamin, New York, 1962).}

\bibitem{jafari08}{S.A.~Jafari, 
%\textit{Introduction to Hubbard model and exact diagonalization}, 
Iranian J.~Phys.~Res.~\textbf{8}, 113 (2008).}

\bibitem{hermanns_tobepublished} S. Hermanns, K. Balzer, and M. Bonitz, to be published.

\bibitem{mctdhf} For a recent overview, see H. Meyer, F. Gatti, and G. Worth,  \textit{Multi-Dimensional Quantum Dynamics}, Wiley-VCH, Weinheim 2010.

\bibitem{hochstuhl_jcp11} D. Hochstuhl, and M. Bonitz, J. Chem. Phys. {\bf 134}, 084106 (2011)

%\bibitem{hochstuhl_pra12} D. Hochstuhl, and M. Bonitz, Phys. Rev. A (2011)

\end{thebibliography}
\end{document}